\documentclass[preprint, double-spaced]{revtex4}
\usepackage{bbm}
\usepackage{color}
\usepackage{graphicx}
\usepackage{mathrsfs}
\usepackage{epsfig,psfrag}
\usepackage{amsmath,amsfonts,amssymb}
\usepackage[dvips, bookmarks, colorlinks=true, plainpages = false, citecolor = blue, linkcolor = blue, urlcolor = blue, filecolor = blue]{hyperref}
\def\be{\begin{equation}}
\def\ee{\end{equation}}
\def\bea{\begin{eqnarray}}
\def\eea{\end{eqnarray}}

\def\smat#1{\left(\begin{matrix}#1\end{matrix}\right)}
\definecolor{mypink1}{rgb}{0.858, 0.188, 0.478}
\definecolor{myred}{rgb}{0.988, 0.088, 0.178}
\definecolor{myblue}{rgb}{0.088, 0.088, 0.978}
\begin{document}
\title{Tight-binding dispersion of the prismatic pentagonal lattice}
\author{Susobhan Paul}\email{suso.phy.paul@gmail.com}
\affiliation{Department of Physics, Scottish Church College, Urquhart Square, Kolkata 700006, India}
\author{Asim Kumar Ghosh} \email{asimkumar96@yahoo.com}
\affiliation {Department of Physics, Jadavpur University, 188 Raja Subodh Chandra Mallik Road, Kolkata 700032, India}

\begin{abstract}
Tight-binding Hamiltonian on the prismatic 
pentagonal lattice is exactly solved to obtain the 
analytic expressions of dispersion relations and eigenvectors. 
This lattice is made of prismatic pentagon which is different from 
Cairo pentagon. 
Six different dispersion relations and total density of states are obtained. 
Dispersion relations are symmetric about the zero energy at a particular point 
in the parameter space. Although a large gap is found for the Cairo pentagonal lattice, 
no gap as well as no Dirac cone is found to appear in the 
tight-binding band structure for this prismatic pentagonal lattice. 
Instead, a pair of van Hove singularities has been identified at two 
different energy values in the band structure. 
\end{abstract}

\maketitle

\section{Introduction}
\label{intro}
\begin{figure}[h]
\centering
\psfrag{aa}{$a$}\psfrag{aaa}{$(1\!-\!\sqrt 3)\,a$}
\psfrag{a1}{$\sqrt3\, a$}
\psfrag{a2}{$5\, a$}
\psfrag{g}{\text{\scriptsize{$\Gamma (0,0)$}}}
\psfrag{m}{\text{\scriptsize{M$\left(\frac{\pi}{\sqrt 3\,a},\frac{\pi}{5\,a}\right)$}}}
\psfrag{x1}{\text{\scriptsize{X$\left(\frac{\pi}{\sqrt 3\,a},0\right)$}}}
\psfrag{x2}{\text{\scriptsize{X$^\prime \left(0,\frac{\pi}{5\,a}\right)$}}}
\psfrag{a}{(a)}\psfrag{b}{(b)}\psfrag{c}{(c)}
\psfrag{d}{(d)}\psfrag{e}{(e)}\psfrag{p}{p}\psfrag{q}{q}
\psfrag{r}{r}\psfrag{s}{s}
\psfrag{t}{$t$}\psfrag{tp}{$\alpha \,t$}
\psfrag{C1}{\text{\scriptsize {\bf C$_1$}}}
\psfrag{C2}{\text{\scriptsize{\bf {\color {white}C$_2$}}}}
\psfrag{C3}{\text{\scriptsize {\bf {\color {white} C$_3$}}}}
\psfrag{B1}{\text{\scriptsize {\bf B$_1$}}}
\psfrag{B2}{\text{\scriptsize{\bf B$_2$}}}
\psfrag{B3}{\text{\scriptsize{\bf B$_3$}}}
\psfrag{kx}{$k_x$}\psfrag{ky}{$k_y$}
\psfrag{d1}{$\bf \delta_1$}\psfrag{d2}{$\bf \delta_2$}
\psfrag{d3}{$\bf \delta_3$}\psfrag{d4}{$\bf \delta_4$}
\psfrag{vd1}{$\bf\delta_1=\sqrt 3\, a\, \hat i$}
\psfrag{vd2}{$\bf\delta_2=a\,\hat j$}
\psfrag{vd3}{$\bf\delta_3=\frac{a}{2}\left(\sqrt 3\,\hat i+  \hat j\right)$}
\psfrag{vd4}{$\bf\delta_4=\frac{a}{2}\left(-\sqrt 3\,\hat i+ \hat j\right)$}
\psfrag{120}{\text{\scriptsize{$120^{^\circ}$}}}
\psfrag{90}{\text{\scriptsize{$90^{^\circ}$}}}
\includegraphics[scale=0.55]{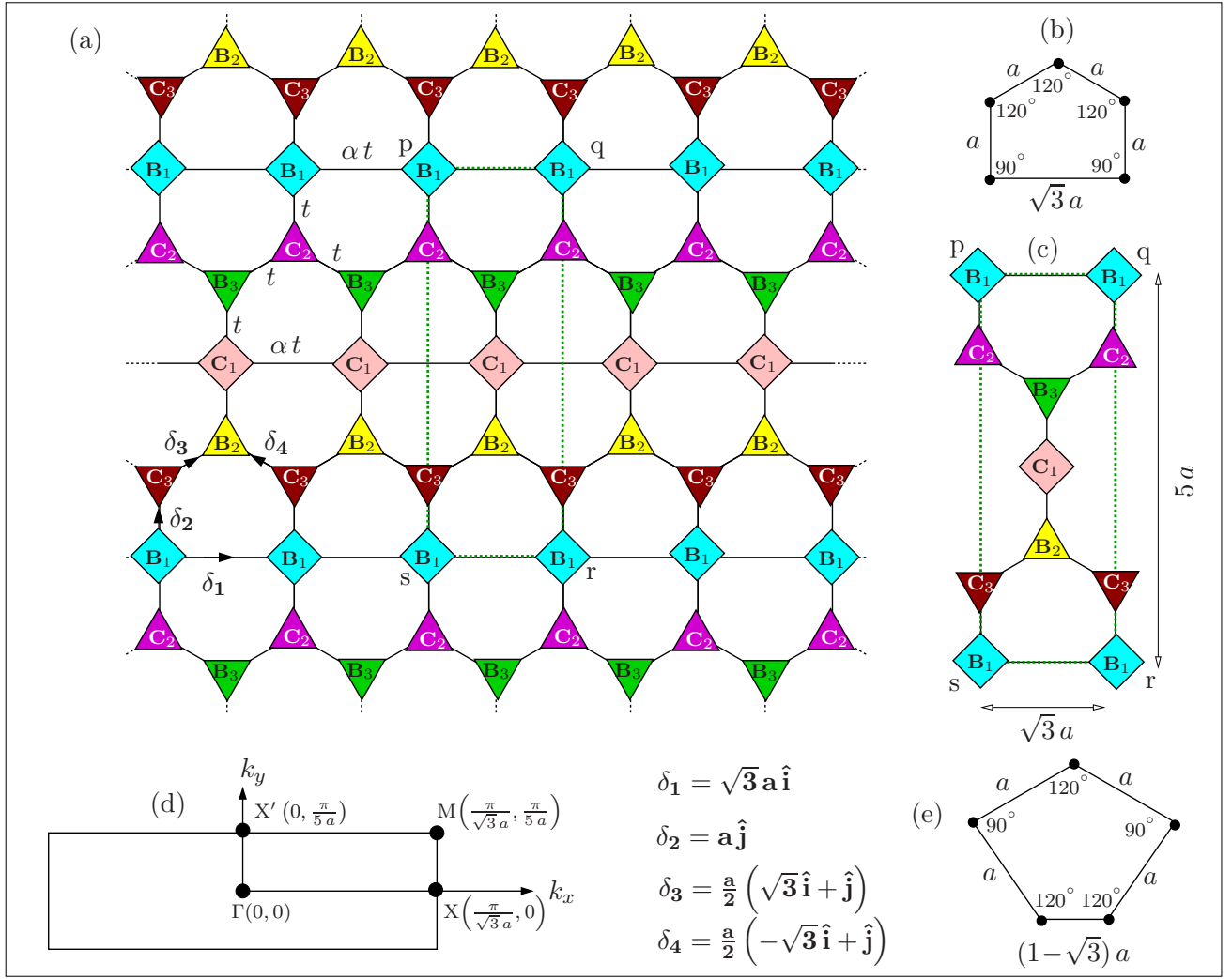}
\caption{(Color online) (a) pentagonal lattice, (b) prismatic pentagon, 
(c) unit cell, (d) first Brillouin zone and (e) Cairo pentagon.} 
\label{lattice}
\end{figure}

Investigations on electronic properties of two-dimensional (2D) models exhibit significant 
growth after the successful realization of graphene, a 2D honeycomb structure of carbons. 
One particular choice of unit cell for honeycomb lattice is a hexagon. 
Although the regular polygons like triangle, square and hexagon 
could fill up a 2D planar surface without 
overlap and leaving any voids, 
it is impossible to do so by using the regular pentagons due to its five-fold symmetry. 
Thus to fill up a planar surface one has to use irregular pentagons those are 
derived from the regular one by changing the 
lengths of the arms as well as the angles between them. 
The examples of irregular pentagons those can fill planar surface are 
prismatic and Cairo pentagons. Structures of prismatic and Cairo 
pentagons are shown in Fig. \ref{lattice} (b) and (e), respectively. 
Electronic properties of the Cairo pentagonal lattice 
have been derived through a number of 
theoretical studies in more recent times. Those results confirm the presence of a 
large gap in its band structure.  
Magnitude of this gap is 3.25 eV which is estimated by 
density functional theory (DFT) on a particular system that is known as penta-graphene (p-graphene).  
P-graphene is essentially a Cairo pentagonal monolayer of carbon atoms \cite{Zhang}. 
DFT analysis on other 
Cairo pentagonal monolayers composed of boron nitride and silver azide  
indicate the presence of a single band-gap as well \cite{Yagmurcukardes}. 
The tight-binding (TB) analysis on p-graphene supports the existence of the same single band-gap whose  
estimated value is very close to the 
previous prediction \cite{Stauber}. 
It has been proposed that a stable carbon allotrope resembling p-graphene 
could be materialized very soon which is entirely composed of Cairo-typed 
carbon pentagons \cite{Zhang}. 
Besides those electronic properties, magnetic properties of 
frustrated antiferromagnetic (AFM) compounds, 
Bi$_2$Fe$_4$O$_9$ \cite{Ressouche}, Bi$_4$Fe$_5$O$_13$F \cite{Abakumov} 
and organic radical crystal, $\alpha$-2,6-Cl$_2$-V 
[=$\alpha$-3-(2,6-dichlorophenyl)-1,5-diphenylverdazyl] \cite{Yamaguchi}
have been reported on the basis of Cairo pentagonal lattice structure. 
An extensive theoretical investigation of magnetic properties on AFM Heisenberg 
Cairo pentagonal lattice could be found in the article \cite{Moessner}.
Ground state phase diagram and magnetization process of the  
spin-1/2 AFM Heisenberg Cairo pentagonal lattice in the presence of 
external magnetic field are available in 
\cite{Sakai1} and \cite{Sakai2}, respectively.  

On the other hand, very few theoretical results are available for 2D lattice 
composed of prismatic pentagons. 
The bond-percolation threshold is determined and the stability of collinear N\'eel-type 
classical ground state for AFM Heisenberg  system 
on this lattice is studied. In addition, ground state entropy and energy are 
estimated for the Ising and AFM Heisenberg models, 
respectively \cite{Waldor,Bhaumik}. 
But, the existence of AFM compounds based on prismatic pentagons are not reported so far. 
Similarly, electronic properties 
of 2D prismatic pentagonal lattice remain unexplored too. 

In this study, we introduce a TB  model on 
the pentagonal lattice that is entirely composed of prismatic 
pentagons (Fig. \ref{lattice} (a)).
The lattice could be derived from the honeycomb lattice in the following way. 
The parent (honeycomb) lattice is elongated vertically in such a manner that 
vertical bonds get doubled in its original length. Additional lattice points 
are introduced in the middle of each vertical bonds. After joining these additional 
points this particular structure of pentagonal lattice is obtained. In the resulting lattice, 
any horizontal layer is made of prismatic  
pentagons by placing them side by side without overlapping and 
leaving any voids. The adjacent layers those are just vertically above 
and below of this one are made of inverted pentagons. The lattice is non-Bravais. 
One may visualize this lattice as composed of  
six interpenetrating rectangular Bravais lattices, 
those are identical to each other. In Fig. \ref{lattice} (a), these sublattices are 
identified by B$_1$, B$_2$, B$_3$, C$_1$, C$_2$ and C$_3$. Those sublattices can be 
divided into two types on the basis of their coordination numbers. 
Coordination numbers of both B$_1$ and C$_1$ are four while they are three for the others. 
The rectangular unit cell of this lattice is shown in Fig. \ref{lattice} (c). 
It effectively contains one lattice point for each of six different sublattices.   
This lattice has the translational invariance of length $\sqrt 3\, a$ along the 
$x$-direction and that of length $5\, a$ along the $y$-direction, 
where $a$ is the length of shorter arms of the pentagon. It also has rotational invariance of 
the angle 180${}^{^\circ}$. 
The first Brillouin zone is shown in Fig. \ref{lattice} (d). 

In this investigation, we have derived the analytic expressions of six different dispersion 
relations exactly along with total density of states (DOS) 
for the prismatic pentagonal lattice. 
The results indicate no gap in the 
band structure for prismatic pentagonal lattice 
in contrast to the case of Cairo pentagonal lattice. 
In addition, a pair of van Hove singularities (VHS) is found to appear at two 
different energy values in the TB band structure of this lattice. 
Therefore, it is expected that electronic properties of 
carbon allotrope based on the prismatic pentagonal lattice
would be different from that of the Cairo pentagonal lattice. 
The TB model and its exact solution are presented in the section \ref{model} 
while the section \ref{conclusion} contains discussion on the results.  

\section{The tight-binding Hamiltonian and dispersion relations}
\label{model}
Considering only nearest-neighbor (NN) hopping, the model 
is described by the tight-binding Hamiltonian,  
\bea
 H\!\!&=&\!\!-t\sum_i\bigg[ b_{1,\bf R_i}^{\dagger} \,c_{3,\bf R_i+\bf \delta_2}+\alpha \,
b_{1,\bf R_i}^{\dagger}\, b_{1,\bf R_i+\bf \delta_1}+b_{1,\bf R_i}^{\dagger}\, c_{2,\bf R_i-\bf \delta_2}
+\,c_{2,\bf R_i+3 \bf \delta_2+\bf \delta_3+\bf \delta_4}^{\dagger}\,
 b_{1,\bf R_i+4 \bf \delta_2+\bf \delta_3+\bf \delta_4}  \nonumber\\   
&+&\!\!c_{2,\bf R_i+3 \bf \delta_2+\bf \delta_3+\bf \delta_4}^{\dagger}\,b_{3,\bf R_i+3 \bf \delta_2+\bf \delta_3}
+c_{2,\bf R_i+3 \bf \delta_2+\bf \delta_3+\bf \delta_4}^{\dagger}\,b_{3,\bf R_i+3 \bf \delta_2+\bf \delta_4}
+b_{3,\bf R_i+3 \bf \delta_2+\bf \delta_3}^{\dagger}\, c_{1,\bf R_i+2 \bf \delta_2+\bf \delta_3}\nonumber\\
&+&\!\!b_{3,\bf R_i+3 \bf \delta_2+\bf \delta_3}^{\dagger}\,c_{2,\bf R_i+3 \bf \delta_2+2 \bf \delta_3} 
+b_{3,\bf R_i+3 \bf \delta_2+\bf \delta_3}^{\dagger}\,c_{2,\bf R_i+3 \bf \delta_2+\bf \delta_3+\bf \delta_4}
+c_{1,\bf R_i+2 \bf \delta_2+\bf \delta_3}^{\dagger}\, b_{2,\bf R_i+\bf \delta_2+\bf \delta_3}\\
&+&\!\!c_{1,\bf R_i+2 \bf \delta_2+\bf \delta_3}^{\dagger}\, b_{3,\bf R_i+3 \bf \delta_2+\bf \delta_3}
+\alpha\,c_{1,\bf R_i+2 \bf \delta_2+\bf \delta_3}^{\dagger}\, c_{1,\bf R_i+\bf \delta_1+2 \bf \delta_2+\bf \delta_3}
+b_{2,\bf R_i+\bf \delta_2+\bf \delta_3}^{\dagger}\,c_{1,\bf R_i+2 \bf \delta_2+\bf \delta_3}\nonumber\\
&+&\!\!b_{2,\bf R_i+\bf \delta_2 +\bf \delta_3}^{\dagger}\,c_{3,\bf R_i+\bf \delta_2}\!+\!
b_{2,\bf R_i+\bf \delta_2 +\bf \delta_3}^{\dagger}\,c_{3,\bf R_i+\bf \delta_2+\bf \delta_3-\bf \delta_4}
\!+\!c_{3,\bf R_i+\bf \delta_2}^{\dagger} b_{1,\bf R_i}\!+\! 
c_{3,\bf R_i+\bf \delta_2}^{\dagger} b_{2,\bf R_i+\bf \delta_2+\bf \delta_3}\!+\!
c_{3,\bf R_i+\bf \delta_2} ^{\dagger}b_{2,\bf R_i+\bf \delta_2+\bf \delta_4}\bigg]\nonumber.
\label{hamilton}
\eea
$t$ and $\alpha\, t$ are the hopping parameters  
across the NN sites having the bonds of lengths $a$ and $\sqrt 3 \,a$, 
respectively (Fig.\ref{lattice} (a)). 
$\bf R_i$ is the Bravais vector for the $i$-th unit cell. 
$b_j^\dagger$ ($c_j^\dagger$) is the creation operator at the $j$-th site 
with respective orbital for an atom on the B (C) sublattice. 
$\bf \delta_i$s are the vectors pointing different adjacent lattice sites within each unit cell. 
Hopping parameters are generally depend 
on the bond lengths in the system. By using the hopping parameter-bond length relationship 
introduced by Harrison, {\em i. e.}, $t(a)\propto 1/a^{ \;2}$, 
value of $\alpha$ is determined, and that is equal to $1/3$ \cite{Harrison}. 
This kind of parametrization for the AFM exchange 
interaction strengths has been adopted before while investigating the stability of 
collinear N\'eel-type classical ground state of Heisenberg Hamiltonian 
on this pentagonal lattice. Interestingly, in that case, the critical value of 
$\alpha$, {\em i. e.}, $\alpha_c$ is also very close to 
$1/3$ above which the classical ground state 
is no longer stable \cite{Bhaumik}. 
We have considered four closer points to each of B$_1$ and C$_1$ sub-lattice sites 
as the NN points despite their different bond lengths. For other sites, number of NN 
points are three and their arrangements are like the honeycomb lattice.  
The orbitals are considered to be the simplest one like the 
$\pi$ electron of graphene for the sites B$_2$, B$_3$, C$_2$ and C$_3$. 
However, no specific orbital structure is assumed for the B$_1$ and C$_1$ lattice sites. 

\begin{figure}[ht]
\centering
\psfrag{a1}{$\alpha=1$}
\psfrag{X}{X}
\psfrag{E}{Energy/$t$}
\psfrag{M}{M}
\psfrag{G}{\text{\scriptsize{$\Gamma$}}}
\psfrag{d}{(d)}
\psfrag{g1}{G$_1$}
\psfrag{kx}{$k_x$}\psfrag{ky}{$k_y$}
\psfrag{e2p}{\text{\scriptsize{\bf $E_2^+$}}}
\psfrag{e1p}{\text{\scriptsize{\bf $E_1^+$}}}
\psfrag{e3p}{\text{\scriptsize{\bf $E_3^+$}}}
\psfrag{e2m}{\text{\scriptsize{\bf $E_2^-$}}}
\psfrag{e1m}{\text{\scriptsize{\bf \color {white} $E_1^-$}}}
\psfrag{e3m}{\text{\scriptsize{\bf \color {white} $E_3^-$}}}
\psfrag{X}{\text{\scriptsize{X}}}
\psfrag{M}{\text{\scriptsize{M}}}
\psfrag{F}{F}
\includegraphics[scale=1.25]{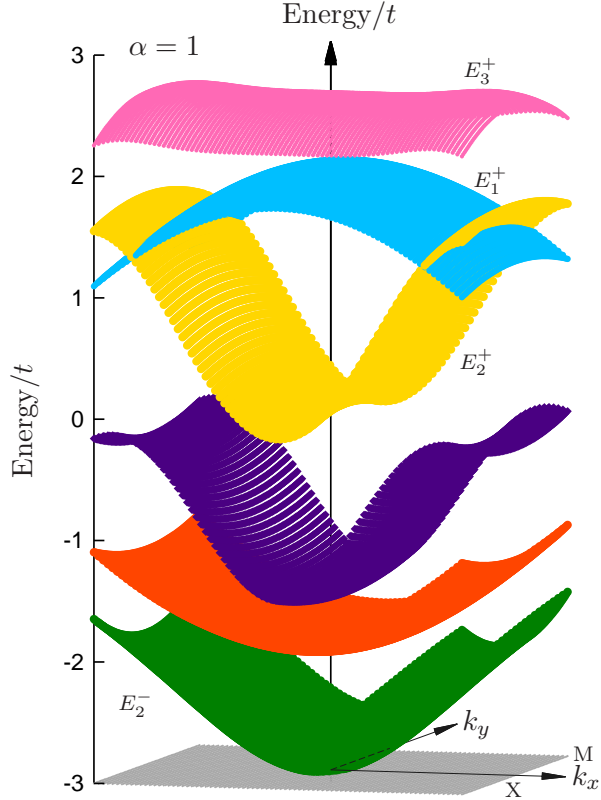}
\caption{(Color online) The dispersion relations,  $E_1^+$, $E_2^+$, $E_3^+$, $E_1^-$, $E_2^-$ and $E_3^-$ 
 for $\alpha=1$ are drawn in different colors within the first Brillouin zone.} 
\label{bands}
\end{figure}

\begin{figure}[h]
\centering
\psfrag{d}{\text{\tiny{\hskip .15cm DOS}}}
\psfrag{X}{\text{\scriptsize{X}}}
\psfrag{M}{\text{\scriptsize{M}}}
\psfrag{Xp}{\text{\scriptsize{X$^\prime$}}}
\psfrag{kx}{$k_x$}\psfrag{ky}{$k_y$}
\psfrag{e2p}{\text{\scriptsize{$E_2^+$}}}
\psfrag{e1p}{\text{\scriptsize{$E_1^+$}}}
\psfrag{e3p}{\text{\scriptsize{$E_3^+$}}}
\psfrag{e2m}{\text{\scriptsize{$E_2^-$}}}
\psfrag{e1m}{\text{\scriptsize{$E_1^-$}}}
\psfrag{e3m}{\text{\scriptsize{$E_3^-$}}}
\includegraphics[scale=1.3]{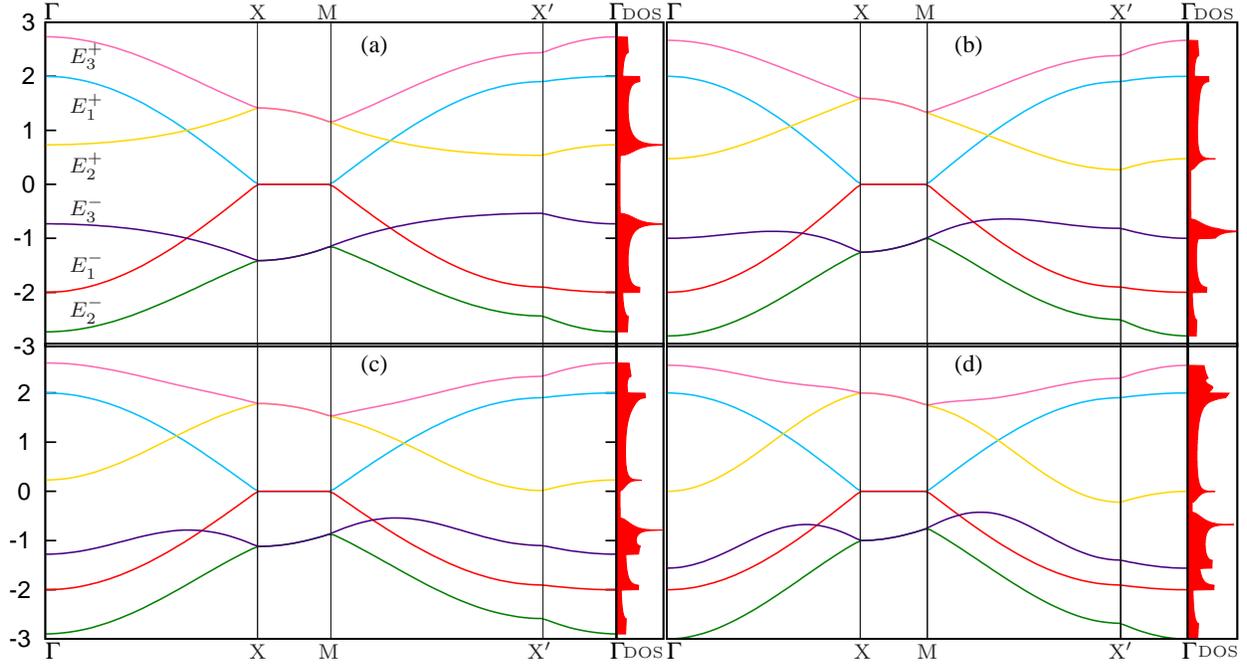}
\caption{(Color online) Dispersion relations, $E_1^+$, $E_2^+$, $E_3^+$, $E_1^-$, $E_2^-$ and $E_3^-$ 
and DOS for $\alpha=0$ (a), $\alpha=1/3$, Harrison point, (b), $\alpha=2/3$ (c) and 
$\alpha=1$ (d) along $q$-path $\Gamma$-X-M-X$^\prime$-$\Gamma$. } 
\label{banddiagrams}
\end{figure}

In order to obtain dispersion relations, the Hamiltonian 
is expressed in $\bf k$-space by transforming the 
$b$ and $c$ operators as 
\[b_{n\,\bf k}=\frac{1}{\sqrt N}\,b_{n,\bf R_i}\,e^{i\bf k\cdot \bf R_i},\quad 
c_{n\,\bf k}=\frac{1}{\sqrt N}\,c_{n,\bf R_i}\,e^{i\bf k\cdot \bf R_i},\quad n=1,2,3,\]
where $N$ is the total number of unit cells in the lattice. 
By introducing the six component vector, 
\[\Psi^\dagger_{\bf k}=\left( b_{1\,\bf k}^\dagger\;\;b_{2\,\bf k}^\dagger\;\;b_{3\,\bf k}^\dagger\;\;
c_{1\,\bf k}^\dagger\; \;c_{2\,\bf k}^\dagger\;\;c_{3\,\bf k}^\dagger\right),\]
$H$ can be written as,  
\be H=-t\sum_{\bf k}\Psi^\dagger_{\bf k}\;H_{\bf k}\;\Psi_{\bf k},\ee
where the $6\times 6$ Hamiltonian in the momentum space 
is decomposed in terms of two $3\times 3$ block Hamiltonians, 
$\Omega_1$ and $\Omega_2$. Therefore, 
\bea
 H_{\bf k}&=&\smat{\Omega_1& \Omega_2\\\Omega_2&\Omega_1},\quad
\Omega_{1}=\smat{X&0&0\\0&0&0\\0&0&0},\quad \Omega_{2}=\smat{0&Y&Y\\Y&0&Z\\Y&Z&0}, \quad {\rm where}
\nonumber \\
X&=&-\alpha\, t \cos{\left(\sqrt 3\, k_x\, a\right)}, \quad Y=-t\cos{\left(k_y \,a\right)}, \quad {\rm and} \quad 
Z=-2\,t\cos{\left(\frac{k_y\, a}{2}\right)}\cos{\left(\frac{\sqrt 3\, k_x\, a}{2}\right)}.
 \nonumber
\eea
The Hamiltonian, $H_{\bf k}$ can be diagonalized by introducing the transfer matrix, 
$T$ such that, 
\[H_{\bf k}=\Phi^\dagger_{\bf k}\;H^d_{\bf k}\;\Phi_{\bf k},\quad \Phi_{\bf k}=T\Psi_{\bf k},\] 
where 
\be
 H^{\rm d}_{\bf k}=\smat{E^+& 0\\ 0&E^-},\quad T=\smat{T_1& T_2\\ T_3& T_4},\quad
E^\pm=\smat{E_1^\pm&0&0\\0&E_2^\pm&0\\0&0&E_3^\pm},
\ee
\be
T_1=\smat{0 & \frac{2Y}{(X-E^+_2)\sqrt{N_+}} & \frac{2Y}{(X-E^+_3)\sqrt{N_+^\prime}} 
 \\- \frac{1}{2} & -\frac{1}{\sqrt {N_+}} & \frac{1}{\sqrt {N_+^\prime}}\\
\frac{1}{2} & -\frac{1}{\sqrt {N_+}} & \frac{1}{\sqrt {N_+^\prime}}},\quad 
T_2=\smat{0 & \frac{2Y}{(X-E^-_2)\sqrt{N_-}} & \frac{2Y}{(X-E^-_3)\sqrt{N_-^\prime}} 
 \\ \frac{1}{2} & -\frac{1}{\sqrt {N_-}} & \frac{1}{\sqrt {N_-^\prime}}\\
-\frac{1}{2} & -\frac{1}{\sqrt {N_-}} & \frac{1}{\sqrt {N_-^\prime}}},\nonumber
\ee
\be
T_3=\smat{0 & \frac{2Y}{(X-E^+_2)\sqrt{N_+}} & \frac{2Y}{(X-E^+_3)\sqrt{N_+^\prime}} 
 \\- \frac{1}{2} & -\frac{1}{\sqrt {N_+}} & \frac{1}{\sqrt {N_+^\prime}}\\
\frac{1}{2} & -\frac{1}{\sqrt {N_+}} & \frac{1}{\sqrt {N_+^\prime}}},\quad
T_4=\smat{0 & \frac{2Y}{(X-E^-_2)\sqrt{N_-}} & \frac{2Y}{(X-E^-_3)\sqrt{N_-^\prime}} 
 \\- \frac{1}{2} & -\frac{1}{\sqrt {N_-}} & \frac{1}{\sqrt {N_-^\prime}}\\
\frac{1}{2} & -\frac{1}{\sqrt {N_-}} & \frac{1}{\sqrt {N_-^\prime}}},\nonumber
\ee
\be
N_\pm = 4\left(1+2\left(\frac{Y}{X-E^\pm_2} \right)^2  \right),\quad 
N_\pm^\prime = 4\left(1+2\left(\frac{Y}{X-E^\pm_3} \right)^2  \right).\nonumber
\ee
$H^{\rm d}_{\bf k}$ is the diagonalized Hamiltonian 
and $\Phi_{\bf k}$ is the eigenvector.
The exact expressions of six dispersion relations,  
$E^\pm_n,\;n=1,2,3$, are written below. 
\bea
E_1^\pm&=&\mp Z, \nonumber \\
E_2^\pm&=&\frac{1}{2}\left(X+Z\pm\sqrt{(X-Z)^2+8Y^2} \right),\nonumber \\
E_3^\pm&=&\frac{1}{2}\left(X-Z\pm\sqrt{(X+Z)^2+8Y^2} \right). \nonumber 
\eea
The dispersion relations, $E^\pm_n,\;n=1,2,3$ for $\alpha=1$ 
have been drawn in different colors 
within the first Brillouin zone (1BZ) and shown in Fig. \ref{bands}. 
The DOS, $g(E)$ could be derived by using the formula,
\[g(E)=\frac{5\sqrt 3}{(2\pi)^2}\int_{\rm 1BZ} d\bf k \!\!\!\sum_{n=1,2,3;\;\gamma=+,-}
\!\!\!\!\!\!\delta(E-E_n^\gamma(\bf k)).\]
Six dispersion relations, $E_1^+$, $E_2^+$, $E_3^+$, $E_1^-$, $E_2^-$ and $E_3^-$ 
along with DOS for $\alpha=0,1/3,2/3$ and 1 along the $q$-path, $\Gamma$-X-M-X$^\prime$-$\Gamma$ 
are drawn in different colors (Fig. \ref{banddiagrams}). 
Dispersion relations $E_1^+$ and $E_1^-$ are always symmetric 
to each other about the zero-energy. 
On the other hand, $E_3^+$ and $E_2^+$ are symmetric to $E_2^-$ and $E_3^-$, 
respectively about the zero-energy only when $\alpha=0$. 
So, DOS is also symmetric about the zero-energy when $\alpha=0$. 
Band gap opens up where DOS vanishes. In this system, 
DOS is always non-zero leading to the appearance of no band gap 
in the TB band structure for any values of $\alpha$. 
No qualitative change in the band structure is observed 
when the amplitudes of hopping parameters across 
$\bf \delta_2$ and $\bf \delta_3$ bonds are made different from 
each other. Only the band-width gets extended when the 
hopping amplitude across the $\bf \delta_{3}$ bond is made 
larger than that of the $\bf \delta_{2}$ bond. 

The appearance of sharp peak in DOS corresponds to the presence of 
VHS in the band structure.
As a result, the nature of DOS in the TB band structure of this 
pentagonal lattice indicates the presence of VHSs at two 
different energy values. 
Appearance of similar pair of VHSs has been observed earlier 
in case of another 2D system, graphene \cite{Neto}.  
Both positions and heights of VHS-peaks vary with $\alpha$. 
For $\alpha=0$, locations of VHS-pair are symmetric about the 
zero-energy in the band structure, which is again similar to the case of graphene. 
Heights of two VHS peaks are the same at this point too. 
Both the symmetries are lost as soon as $\alpha\neq 0$. 
With the increase of $\alpha$, separation between two VHSs gets reduced. 
Height of the upper VHS peak becomes shorter and it moves toward bottom  
with the increase of $\alpha$. On the other hand, 
height of the lower VHS peak increases with the increase of $\alpha$ and 
it shifts toward bottom when $0\!<\!\alpha\!<\!1/2$. But the opposite behavior  
is observed in the same lower VHS peak when $1/2\!<\!\alpha\!<\!1$. 
VHS could be detected and characterized with the help of 
scanning-tunneling, optical and Raman spectroscopies whenever  
it appears very close to the Fermi energy ($E_{\rm F}$). 
In graphene, positions of VHSs are symmetric about $E_{\rm F}$ and they 
emerge at far away from $E_{\rm F}$, which 
make them very difficult to observe \cite{Neto}.  
Relative positions of VHSs could be changed by twisting the graphene layers, 
which in turn alter the values of hopping parameters \cite{Harrison}. 
Observation of low-energy VHSs by using scanning tunneling spectroscopy in the 
twisted graphene layers has been reported in the article \cite{Andrei}. 
Similarly, in case of prismatic pentagonal lattice, positions as well as heights 
of VHS-peaks could be varied by controlling the value of $\alpha$. 
VHS largely affects the optical and electrical properties of the materials. 
For example, optical absorption spectra of the twisted bilayer graphene 
are found to exhibit multiple peaks, which corresponds to the 
presence of multiple VHSs in its band structure \cite{Koshino}.
However, in two-dimensional systems, electronic instabilities occur  
when VHS appears in the vicinity of $E_{\rm F}$ leading to the 
emergence of several phases of matter like superconductivity \cite{Kohn}, 
ferromagnetism \cite{Fleck} and charge density waves \cite{Rice}.
\section{Conclusion}
\label{conclusion}
In this study, we have clearly predicted the existence of no band-gap in the TB analysis on the prismatic 
pentagon lattice in contrast to the previous observation of a 
single band-gap for the Cairo pentagonal lattice. 
In addition, 
a pair of VHSs appears in the band structure at two different energies. 
Similar feature is observed before in graphene. 
Both positions and heights 
of the VHS-peaks are found to vary with the values of $\alpha$. 
Like the Cairo pentagonal lattice, no Dirac-cone is found. It would be noted that 
Dirac-cones are omnipresent in the 2D lattices with the 
coordination number of three, such as, honeycomb (graphene), 1/2-depleted 
(T-graphene) and 1/5-depleted square lattices \cite{Liu}. 
However, in these pentagonal lattices coordination numbers are either three or four depending on 
the position of lattice sites. Recently, innovation of another class of 2D materials, such as  
p-graphene \cite{Zhang} and heptagraphene \cite{Lopez},  
resembling the graphene structure 
are gaining interests in which the basic structure of those lattices are generated by using the 
irregular pentagons and heptagons, respectively. This trend is due to the appreciation of 
exotic electronic properties observed in graphene. 
Although, p-graphene is made of only carbon atoms, the unit cell of heptagraphene is 
composed of ten carbon and four hydrogen atoms. Again, 
triggered by the success of graphene physics, realization  
 of new carbon allotropes are gaining considerable impetus.  
Possibility of stable p-graphene based on the Cairo pentagon may be realized very soon.  
Therefore, it is expected that 
another stable p-graphene based on the prismatic pentagonal lattice  
would be materialized in near future. 
Apart from the pure Cairo and prismatic pentagonal lattices, a class of 2D lattice structures 
made of several combinations of Cairo and prismatic pentagones 
could be generated which might give rise to novel electronic properties as well. 
\section{acknowledgements}
 Authors are grateful to Prof. Indrani Bose for bringing 
the pentagonal lattice model to their notice. 
AKG acknowledges a BRNS-sanctioned 
research project, no. 37(3)/14/16/2015, India. 

\end{document}